\documentclass[12pt]{article}
\usepackage{geometry}
\usepackage[utf8]{inputenc} 
\usepackage[T1]{fontenc}   

\usepackage{float}
\usepackage{array}
\usepackage{setspace}
\usepackage{amsmath}

\usepackage[numbers]{natbib}

\usepackage{siunitx}
\usepackage{amssymb}
\usepackage{tikz}
\usetikzlibrary{shapes, positioning}
\usepackage{url}

\title{Towards Scalable Insect Monitoring: Ultra-Lightweight CNNs as On-Device Triggers for Insect Camera Traps}

\author{
    Ross Gardiner \textsuperscript{1}, Sareh Rowlands\textsuperscript{2}, Benno I. Simmons\textsuperscript{3} \\[1ex]
    \textsuperscript{1} Environmental Intelligence CDT, University of Exeter, UK\\
    \textsuperscript{2}Institute for Data Science and AI, University of Exeter, UK \\
    \textsuperscript{3} Centre for Ecology and Conservation, University of Exeter, UK
}

\date{} 
\begin{document}

\maketitle


\section*{Abstract}
\begin{enumerate}
    \item Camera traps, combined with AI, have emerged as a way to achieve automated, scalable biodiversity monitoring. However, the passive infrared (PIR) sensors that typically trigger camera traps are poorly suited for detecting small, fast-moving ectotherms such as insects. Insects comprise over half of all animal species and are key components of ecosystems and agriculture. The need for an appropriate and scalable insect camera trap is critical in the wake of concerning reports of declines in insect populations.
    \item This study proposes an alternative to the PIR trigger: ultra-lightweight convolutional neural networks running on low-powered hardware to detect insects in a continuous stream of captured images. We train a suite of models to distinguish insect images from backgrounds. Our design achieves zero latency between trigger and image capture. 
    \item Our models are rigorously tested and achieve high accuracy ranging from 91.8\% to 96.4\% AUROC on test data and 58.8\% to 87.2\% AUROC on field data from distributions unseen during training. The high specificity of our models ensures minimal saving of false positive images, maximising deployment storage efficiency. High recall scores indicate a minimal false negative rate, maximising insect detection. Further analysis using saliency maps shows the learned representation of our models to be robust, with low reliance on spurious background features. Our method is also shown to operate deployed on off-the-shelf, low-powered microcontroller units, consuming a maximum power draw of less than 300mW. This paves the way for scalable systems with longer deployment times.
    \item Overall, we fully define the properties of a successful trigger for camera traps and show how lightweight AI models, made bespoke for efficient hardware can be realised with a specific focus on insect ectotherms. We provide these models to the community alongside a complete codebase for future modifications and we demonstrate how they can be deployed on an example ESP32-S3 microcontroller platform. This step potentiates a major advancement in ectotherm camera traps and insect monitoring.

\end{enumerate}

\section*{Data/Code}

All datasets used for training and testing
along with the code-base and all realised models will uploaded to the Data Dryad platform upon acceptance at: \url{https://datadryad.org/stash/dataset/10.5061/dryad.p5hqbzkz7}. \\
The code-base is attached to this submission for peer-review. Ecto-Trigger project documentation, codebase and user guidance is available at: <REDACTED\_FOR\_REVIEW> 

\section*{Keywords}
Artificial Intelligence, Computational Entomology, Conservation Technology, Camera Traps, Insect Declines, TinyML, Biodiversity, Insects

\section{Introduction}\label{sec:intro}
\doublespacing
Insects are key components of all terrestrial ecosystems, comprising over half of all described animal species \citep{80percent-insects}; they are essential to agricultural systems for producing crops and feedstock sustainably, and they play a major role in health as disease vectors affecting humans and animals \citep{insect-disease-vector}. Declines in insect biomass and diversity changes have received attention in recent years \citep{insect-decline-meta-analysis, death-by-thousand-cuts, 75percent-decline}. Traditional insect monitoring fieldwork methodologies — such as Pan Traps, Pitfalls, and Malaise Traps — scale poorly, due to high labour and time costs \citep{standard-insect-monitoring}. Development of scalable survey techniques is critical for global-scale monitoring of the extent and steepness of declines \citep{scalable-insect-monitoring, standard-insect-monitoring}. Technological advancements in electronics, remote cameras and artificial intelligence offer a route to meet this challenge \citep{synergistic-future, deep-learning-will-transform-entomology}. 

Camera traps have emerged as a popular method to survey populations of large mammals and birds \citep{scaling-camera-traps}. Due to their autonomous nature, they enable continuous (24/7) monitoring across large spatial scales utilising many cameras at low cost. The large quantities of image/video data produced by camera traps have necessitated the development of AI methods that can automatically identify species present in each image/video \citep{noro-pnas, whytock}. Integration of camera traps and AI, handling data processing, completes an end-to-end architecture suitable for fully automated, and thus globally scalable biodiversity monitoring. However, due to remote deployments where access to energy sources and data transfer capabilities are limited, this scalability is constrained by two critical factors: power consumption and data volume.

Integral to these constraints is the trigger mechanism, a signal that determines when a camera trap captures images/video. Designing an effective trigger is challenging as it must optimise each of the following aspects: 
\begin{enumerate}

  \item \textbf{Storage Efficiency}: Since a trigger always signals data capture, it governs the quantity of data which a system gathers, directly impacting deployment time as full SD cards require manual changing. 
  \item  \textbf{Responsiveness}: Since a trigger is the only signal for data capture, it must consistently activate when the target is present. A trigger which is always activating is maximally responsive but will also capture the maximum amount of false positives (“blank” images/video), compromising storage efficiency.
  \item \textbf{Latency}: We define this as the time taken between trigger activation and image capture. Minimal latency is required to ensure that fast-moving animals are still within camera field-of-view (FOV) when images/video are captured. A trigger system with excellent responsiveness can still save blanks if latency is high. 
  \item \textbf{Energy}: Triggers are in constant operation and thus can have the largest impact on system power consumption, affecting the scalability of camera trapping surveys via battery life depletion and expense of additional energy systems (e.g. larger batteries or solar panels). 
   
\end{enumerate}
Traditional camera traps have typically balanced these requirements via the use of passive infrared (PIR) sensors which signal to record data only when sudden changes in surface temperature are detected \citep{pir-misconceptions} and consume negligible power otherwise. They can be deployed in remote locations for months at a time, maximising the chance to observe rare or elusive species and minimising human effort. However, PIR triggers are not appropriate for invertebrates \citep{halt-hobbs}, due to a lack of detectable body heat. When PIR-triggered camera traps have been deployed for insect monitoring trials, they suffer from poor storage efficiency, resulting in many ``blank'' images/videos \citep{moth-pir} and poor responsiveness compared with scheduled image capture \citep{time-lapse-insects}. Consequently, the design of insect-focussed monitoring devices cannot rely on a conventional triggering method. 

There are some alternatives which negate the requirement for a trigger. \citeauthor{integrated-pest-monitoring-cameras} surveys camera systems for integrated pest monitoring. These systems typically attract species (e.g. pheromone lures) and capture them fatally (e.g. sticky paper or liquid). Often cameras then take time-lapsed footage giving temporal resolution \citep{integrated-pest-monitoring-cameras}. Whilst no trigger is required, the ethics of this method are not appropriate to scaling conservation efforts \citep{ethics-lethal}. Conversely, time-lapsed footage without lethal means disregards storage efficiency. Event-camera based systems offer low power consumption \citep{gebauer2024towards}, but these are currently prohibited by cost. Options based only on passive acoustics can also operate with low power consumption \citep{buzzometer} and prior work has shown that wing-beat harmonics can discern a modest number of species from background noise \citep{9-month-lidar}. While audio provides a data-efficient, low-power route, image data is crucial for developing and applying existing AI models with broad taxonomic coverage \citep{stevens2024bioclip, insect-foundation, Hierarchical-insects, august-metadata}. Thus, the search for an effective insect camera-based trap remains the most active.

The plant–insect interactions camera trap (PICT), without a trigger mechanism, continuously records video to disk and can achieve low power draw utilising optimisations provided by the Raspberry Pi system \citep{pict}, it can run for a few days before storage is depleted. More recent alternatives opt to integrate AI capabilities into devices and perform analysis in the field to reduce the empty image storage burden \citep{Insect-detect, darras, amt, bjerge-jetson}. Such AI pipelines developed for insect camera traps have followed trends in traditional camera trap AI development; object detection models initially find areas of the image which contain insects and classifiers determine species labels. A problem with this paradigm is that the compute power required for these object detectors is more than the PICT; with consequent power consumption on the order of several 1.3-12.5W for each system (further details in Table \ref{tab:look-at-me}). Their advantage is continuous recording with the recommended addition of continuous energy via solar panels. While effective, this makes systems more expensive due to the cost (e.g. of charge control boards and solar panels) and less portable due to deployment requiring continuous strong sunlight for energy harvesting. A trigger to accurately determine data volume only at low energy cost remains a critical ingredient.


Here, we explore how all aspects of the trigger problem may be tackled using image-based ultra-lightweight AI designed to execute on microcontroller units (MCUs) \emph{at the edge}, meaning deployed on platforms in the field where power and data volume limitations are significant. Our design recognises (1) that image data are required to achieve high levels of species granularity; (2) that a trigger must be storage efficient, responsive, and operate with minimal latency and (3) that the trigger should suit hardware platforms which are low-powered, in this case we target MCUs. Our trigger allows a continuously captured stream of images to be processed and searches for those with insect features as learned by a convolutional neural network (CNN). Via carefully constructed accuracy measures, we measure the storage efficiency and responsiveness of the trigger. We also show the trigger operating on the latest power-efficient microcontroller hardware to measure its power consumption and quantify quantisation error. Further, to build trust in our method, we extensively evaluate our models on test datasets, example images from field trials and via analysis with saliency maps to explain the representations learned by the models. The result is a suite of AI models with especially low computational burden, high accuracy and trustworthiness. We provide all model weights, extensive software documentation and guidance for others to develop their own variations for ectotherm-focussed camera traps.

\section{Materials and Methods}\label{sec:methods}

\begin{figure}[H]
    \centering
    \includegraphics[width=0.8\linewidth]{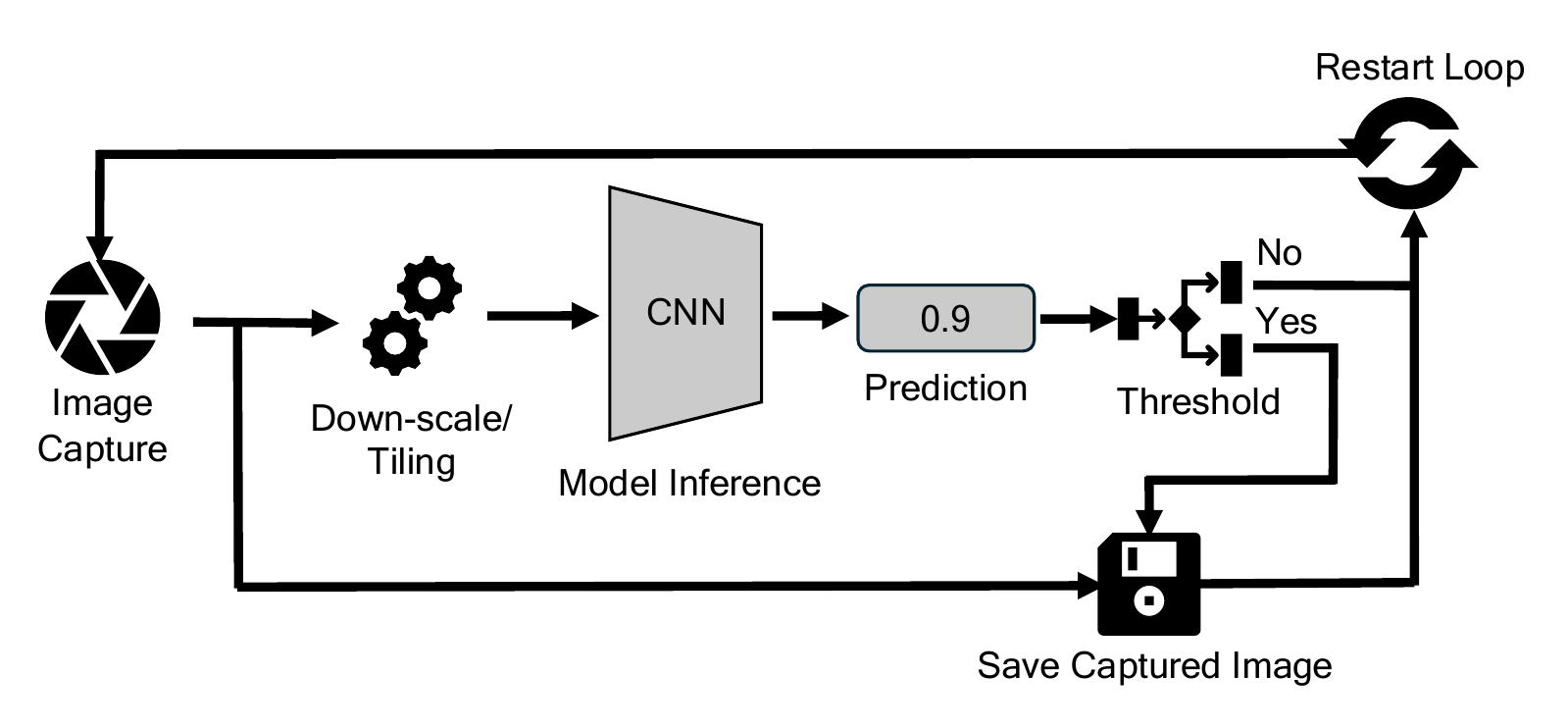}
    \caption{Software pipeline showcasing how the trigger (CNN, binary classifier) can be built into programmes on deployed devices. Initially, images are captured and a preprocessed variant of the image is fed into the model for prediction. Logic thresholds the prediction to warrant saving the original image, this loop then repeats.}
    \label{fig:pipeline}
\end{figure}

Creating a CNN model for execution on MCUs has many challenges as these so-called ``TinyML" algorithms are severely constrained by the computing capabilities of host devices \citep{tinyml}. This is because (1) the execution memory available on the device limits the number of model parameters which can be used and (2) low clock speeds of MCUs increase the computation time for model inference. Initially, object detection architectures seem attractive; they excel at finding small animals in images \citep{embedded-camera-trap-images} and have been widely used for camera trapping workflows. However, object detection models require a significant number of parameters, regularly increasing tenfold or more when compared with more simple classifier models \citep{mobilenet}, making them ill-suited to microcontroller deployment. Thus, we take inspiration from the ``Visual Wake Words'' concept \citep{visualwakewords} which reliably identifies human-containing images using a MobileNet classifier with far fewer parameters than MobileNet object detectors. Once a model is found, it can be integrated into microcontroller firmware to continuously capture images and determine if they contain an insect. This workflow, shown by Figure \ref{fig:pipeline}, eliminates latency between trigger and image capture by making decisions retrospectively, eliminating the risk of blank images caused by slow trigger speeds. The original image can also be saved at full resolution for downstream processing after a decision is made. The rate at which image capture can be achieved is determined by the execution latency of the model, faster models ensure higher detection probability of fast animals moving through the camera field-of-view (FOV).

Following the ethos efficient models,we use the MobileNetv2 architecture, which incorporates features that reduce computation, such as depthwise separable convolutions, residual bottleneck layers, and the ReLU6 activation function, which is tolerant to low precision \citep{mobilenetv2}. MobileNetv2 also includes an $\alpha$ parameter that adjusts the network's width, enabling a trade-off between model complexity and computational efficiency; $\alpha$ values below 1.0 reduce the network's width and computational burden \citep{mobilenet}. For a full description of the MobileNetv2 architecture, we refer the reader to the respective MobileNet and MobileNetv2 papers. We selected modest values $\alpha$ values of 0.1 and 0.35 as part of a heuristic approach to explore suitable configurations for deployment on resource-constrained microcontroller hardware. We experimented with removing input colour channels to reduce computations per inference. We also simplify the learning problem to a basic form of binary classification, using a single output neuron to indicate insect presence (\verb|1.0|) or absence (\verb|0.0|) in the input. To further save computation, our models use modest input sizes, the smallest of which is 96x96x1 and the largest of which is 120x160x3. This results in a small number of parameters for each model, as shown by Table \ref{tab:models}.

Using a smaller input size for the models has several down-stream implications: (1) image down-scaling may be required to input an image to a model, potentially reducing detection probability for small objects in high-resolution images with a wide FOV, as is common for insect camera trap field datasets \citep{Insect-detect, ecostack, pollinator-detect, ami-dataset}. When resized, these objects lose excessive detail, making them difficult to discern (Figure S3). To address this, we propose a “tiling” approach, which allows the model to repeatedly process smaller sections of each image at higher resolution. A detailed description of this method is provided in Section \ref{sec:tiling}. (2) Implementing a tiling strategy affects the number of full-frame images the system can process per second as each tile requires further computation time. (3) If a tiling strategy is not used, field deployment setups should ensure the subject occupies a larger portion of the image. This approach is suitable for scenarios like pollination studies or lure-based setups, where insects are expected in a specific area. Positioning the camera closer reduces the field of view, aligning with the model's capabilities. Although smaller fields of view limit the sampling area, this can be balanced by scalability, such as longer deployments or deploying more systems \citep{darras}.

One final step we have taken to ensure these models can run on MCU platforms is using quantisation. This technique reduces the precision of a neural network’s internal data (i.e. weights and activations) into representations requiring a lower number of computational bits to represent them. Quantisation is commonly used for neural network deployments to microcontrollers as their computation architectures are typically optimised for operations on 8-bit integers. There are many flavours of quantisation, we utilise post-training quantisation, which turns our floating point weights into 8-bit signed integers. This is performed using the tools provided by TensorFlow, and follows the protocol outlined by \cite{tflite-paper}. Critically, we note that the compression of the model data incurs an accuracy penalty and we monitor this throughout our evaluation.

Below are details of our method for training and evaluation. Specialised datasets for the binary insect classification task are also presented. Code for training and evaluation work was written with iterative consultation from the ChatGPT LLM, version 4o, and checked extensively by the lead author.



\begin{table}[H]
    \centering
    \begin{tabular}{c|c|c|c}
         Model name & $\alpha$ & Input shape (HxWxC) & Nr. Parameters \\
         1 & 0.1 & (96x96x1) & 94,449\\
         2 & 0.1 & (96x96x3) & 94,593\\
         3 & 0.1 & (120x160x1) & 94,449\\
         4 & 0.1 & (120x160x3) & 94,593 \\
         5 & 0.35 & (96x96x1) & 411,201 \\
         6 & 0.35 & (96x96x3) & 411,489 \\
         7 & 0.35 & (120x160x1) & 411,201 \\
         8 & 0.35 & (120x160x3) & 411,489\\
         
    \end{tabular}
    \caption{Model variations for binary classification, showing variations on input size and the $\alpha$ parameter for each model labelled 1-8. The number of model parameters as reported by Keras is shown in the rightmost column and range from 94,449-441,489. Models were instantiated with the Adam optimiser \citep{adam}, a fixed learning rate, $\mu$ = \texttt{0.001}.}
    \label{tab:models}
\end{table}
\subsection{Datasets}\label{subsec:datasets}
Selection of an appropriate training dataset was driven by three main factors: (1) the diversity of insect species and inclusion of varied background examples, (2) inclusion insect images captured in the field to improve model transfer to real-world field conditions, (3) the quantity and quality of insect images.

We explore a variety of source datasets for insect classification and detection. The AMI-GBIF dataset contains over 2.5 million images of 5,364 species \citep{ami-dataset}, it depicts insect images from citizen science platforms and from museum specimens. Thus, the context of insects is varied, with some having diverse backgrounds in-situ and others with homogenous backgrounds such as backboards from museum collections. Other datasets contain insects images observed by automated insect camera traps \citep{ecostack, Insect-detect, pollinator-detect}. Leveraging these is judged critical to improve the transferability of learned representations to automated field scenarios but they are limited by small sample sizes featuring only a few species and images with homogenous backgrounds. We elected to construct a dataset comprising images from the field combined with varied examples from the iNaturalist 2017 challenge dataset \citep{inaturalist}. 

We select the iNaturalist 2017 dataset due to its broad diversity encompassing 1021 insect species. We use those annotated with the 125,679 additional bounding boxes (``iNat2017-insecta''). This dataset includes a wide range of bounding box sizes, many small, this is important for training the model to recognise different subject sizes. As negative examples, we also include the iNaturalist 2017 \emph{plantae} images as these contain a variety of background foliage (``iNat2017-plantae''). Class imbalance is known to bias learned representations, so we carefully balance the number of background and insect images with an equal number of examples from each source. iNat2017-plantae images are randomly sub-sampled, their quantity is equal to iNat2017-insecta. Images containing insects were assigned the label \verb|1| and \verb|0| otherwise.

To add field deployment data, we supplement our dataset with images captured by existing insect camera traps. We use data from \citet{ecostack} (``Ecostack'') and \citet{Insect-detect} (``Insect-Detect''). These make up a minority portion of the training and test dataset and they cover 9 and 5 insect classes respectively; labels are also supplied in the bounding box format denoting the spatial location of an insect in each image. The overall training and test data mix is shown by Table \ref{tab:data}.

\subsection{Tiling}\label{sec:tiling}

Our models require small input sizes, making resizing a challenge. Ecostack and Insect-Detect feature high resolution images where insects occupy only a small portion of the frame, thus, simply downscaling them is impractical for our models as important subject details are lost (Figure S3). To address this, we preprocess these images by splitting them into a grid of "tiles", similar to a jigsaw puzzle, ensuring the tile's size is a factor of $2^n$, relative to the original image size (Figures S1 and S2 show tile examples). To convert their labels from bounding boxes to binary classification, we assign a label of \texttt{1} to a tiled image if it contains a bounding box with more than 50\% of its area prior to tiling within the tiled image. Images with no overlapping bounding boxes are labelled as \texttt{0}. Images containing overlapping bounding boxes covering less than 50\% of the bounding box are not included in our dataset. Discarding these images encourages the model to focus on clear, complete insect examples, avoiding confusion from challenging cases with minimal insect content. A disadvantage of this approach is that the training data are made less various, this may reduce generalisation. A 50\% threshold is selected to balance these trade-offs. All bounding box class labels are treated as insect instances, except for the \texttt{shadow} class in Insect-Detect, which is treated as background. To balance the insect and background images, the required number of insect images (Table \ref{tab:data}) are selected, and background tiles are randomly sampled to match this number or exhaust those available.



\begin{table}[H]
\resizebox{\textwidth}{!}{%
\begin{tabular}{l|c|c|c|c}
Source &
  \begin{tabular}[c]{@{}c@{}}Insect Images\\ (train)\end{tabular} &
  \begin{tabular}[c]{@{}c@{}}Background Images\\ (train)\end{tabular} &
  \begin{tabular}[c]{@{}c@{}}Insect Images\\ (val)\end{tabular} &
  \begin{tabular}[c]{@{}c@{}}Background Images\\ (val)\end{tabular} \\ \hline
iNat2017-insecta           & 97524  & -      & 16732 & -     \\
iNat2017-plantae           & -      & 97524  & -     & 16732 \\
Ecostack n=2       & 11000  & 11000  & 1840  & 1840  \\
Ecostack n=3       & 1981   & 1726   & 3215  & 2989  \\
Insect-Detect n=2  & 1328   & 1328   & 190   & 190   \\
Insect-Detect n=3  & 1456   & 1151   & 185   & 154   \\ \hline
Total                      & 113289 & 112729 & 22162 & 21905
\end{tabular}%
}

\caption{Image quantities in training and test datasets for binary classification including breakdown of positive and negative sample quantities across iNat2017-insecta, iNat2017-plantae, For the test set, examples from each dataset source's respective validation set are used. A total of 135,451 images containing insects and 134,634 images containing backgrounds results in a train/test ratio of 84\%/16\%. Image samples from  Ecostack and Insect-Detect are tiled such that the resulting images are one quarter and one eighths of the size of the original images; their tiling parameters, n, are set to 2 and 3 respectively. }
\label{tab:data}
\end{table}

\subsection{Training}\label{subsec:training}
The training was undertaken using a single NVIDIA V100 GPU for each model, utilising computational resources from the JADE2 HPC facility and completed after one week of computation time.

Each model was instantiated in the Keras programming environment \citep{keras}, using default implementations and hyperparameters listed in Table \ref{tab:models}. Transfer learning accelerates model training and improves performance by leveraging knowledge from related tasks; for example, a model pre-trained on human images may gain better performance from further training to distinguish individuals versus training from a randomised start point. To leverage these advantages, models were first pre-trained for 100 epochs on the iNat2021-mini dataset with randomly initialised weights. iNat2021-mini is a variation on the iNaturalist 2021 challenge dataset which is miniaturised by reducing the number of images for each of the 10,000 classes to 10 \citep{inat2021}. Classes are species spread across the Tree of Life and this helps the models learn generic features for natural images. During this stage, a classification head of size 10,000 was added, which refers to the final layer responsible for mapping the learned features to the 10,000 output classes. Categorical cross-entropy loss was used to facilitate learning, with training conducted using a mini-batch size of \texttt{64} and NVIDIA CUDA acceleration for TensorFlow.

Each image was resized to fit the input layer using bilinear interpolation. For greyscale images, RGB inputs were converted using the \texttt{cvtColor} function available from the OpenCV library. As is common in deep learning, we apply image augmentations to further diversify our training data. Augmentations are (often) basic operations which modify images to provide new examples to the model. We have applied an operation which flips images around the vertical axes and executes with a likelihood of 50\%; the training data are diversified with mirror images. Regularisation is another technique which improves model performance by artificially making training more challenging. We added this via a dropout layer, which is a component that randomly deactivates neuronal connections during training, with a probability of 50\% \citep{dropout}. Our dropout layer was added to the penultimate network layer, before the head.

After pre-training, the classification head was replaced with a single output neuron for binary classification. The main training comprised a further 10 epochs on our dataset, using the same method as above and replacing the categorical cross-entropy loss function with binary cross-entropy. After training, each model has had its weights quantised such that they can be represented with an \verb|int8| (signed 8-bit integer). We utilised the TensorFlow Lite for Microcontrollers software package to quantise our models \citep{tflite-micro}. 

\subsection{Evaluation}\label{subsec:eval}
We conduct a comprehensive evaluation of our models to assess their suitability as camera trap triggers, focussing on the characteristics outlined in the Introduction. First, model accuracy metrics are used to evaluate the storage governance and trigger responsiveness, then saliency maps are used to evaluate the model's insights, finally an example MCU deployment quantifies power consumption.

\subsection{Accuracy Metrics}

There are many correctness metrics for a binary predictive test, each uses a combination of the following elements: (1) True Positives (TP), the number of positive and correct predictions; (2) True Negatives (TN), the number of negative and correct predictions; (3) False Positives (FP), the number of positive and incorrect predictions; and (4) False Negatives (FN), the number of negative and incorrect predictions. A commonly used metric is \textit{accuracy}, which is the fraction of correct predictions (TP + TN) out of all predictions (TP + TN + FP + FN), this measure gives a good indication of overall performance, but, it does not differentiate the impact of TP versus TN cases. Further metrics include the \textit{precision}, which is the number TP out of all positive predictions (TP + FP) - this evaluates the correctness of positive results only. Another is \textit{recall} which evaluates TP as a fraction of the total positive examples that were available (TP + FN) - this evaluates the ability to find all positive examples and thus models the trigger responsiveness. The \textit{F1 score} is an average of precision and recall. While useful, this measure omits the value of TN. This is significant, as negatives may be very high in a scenario where insects visits are rare, thus the ability to identify negative cases is important to understand the trigger's storage governance \citep{motion-vectors}. To address this, we also consider \textit{specificity} which can be thought of as the recall of negatives; it is the portion of TN out of all available negatives (TN + FP). 

A further challenge is considering where to place the threshold at the model output before a positive prediction can be declared. Different ecological scenarios may have different requirements. In a situation where one cannot afford to miss an observation, this threshold may be set low (i.e. 0.1) to ensure the output is more readily positive at the expense of a greater number of FP. On the contrary, in a scenario where storage governance is paramount, the threshold may be set higher (i.e. 0.9) at the expense of FN. As these use-cases compete, we summarise them using a receiver operating characteristic (ROC) curve, which plots the recall and specificity at various output thresholds \citep{ROC-curves}. We also distill this curve to a single value by computing the area under the ROC curve (AUROC), appropriately capturing overall performance across all thresholds and allowing models to be easily compared across various configurations and datasets. A model with a 1.0 AUROC value is perfect, one with 0.5 has no discriminative power and one with an AUROC value of 0.0 perfectly miss-labels all examples at all thresholds. We recognise that many other works in this area investigate accuracy, precision, and F1 scores so we also provide these on the withheld test set at an output threshold of 50\%. 

Using this comparison method, also compute the AUROC for both Ecostack and Insect-Detect and for two field insect camera trap datasets unseen in training. We choose \citet{pollinator-detect}'s dataset of pollinator images (Pollinator-Detect) and the camera trap images from the AMI Dataset (AMI-CT) \citep{ami-dataset}. These data are tiled in the same manner as described in Section \ref{sec:tiling}, resulting in a further challenging test from realistic unseen field data.

\subsection{Saliency Maps}

Next, we investigate the legitimacy of the features learned by the model. A common problem in supervised learning is overfitting to training data characteristics that fail to generalise in new domains. For example, \citeauthor{camera-trap-backgrounds} found classification performance to drastically reduce when the same animal is shown with different backgrounds \citep{camera-trap-backgrounds}. To build trust in the learnt features, we endeavour to determine the most impactful image regions using saliency maps. Saliency maps graphically show the output gradient with respect to the input image, indicating how sensitive the model's prediction is to changes within pixel regions. While there are many ways to create saliency maps \citep{molnar2022}, we chose the "Vanilla Gradient" method because it is the simplest. This method calculates how each region in an image influences the model's prediction by measuring how a small change in each pixel affects the output value. The result is a heatmap, the same size as the input image, where brighter areas show the pixels with steepest output gradient.

\begin{figure}[H]
    \centering
    \includegraphics[width=0.8\linewidth]{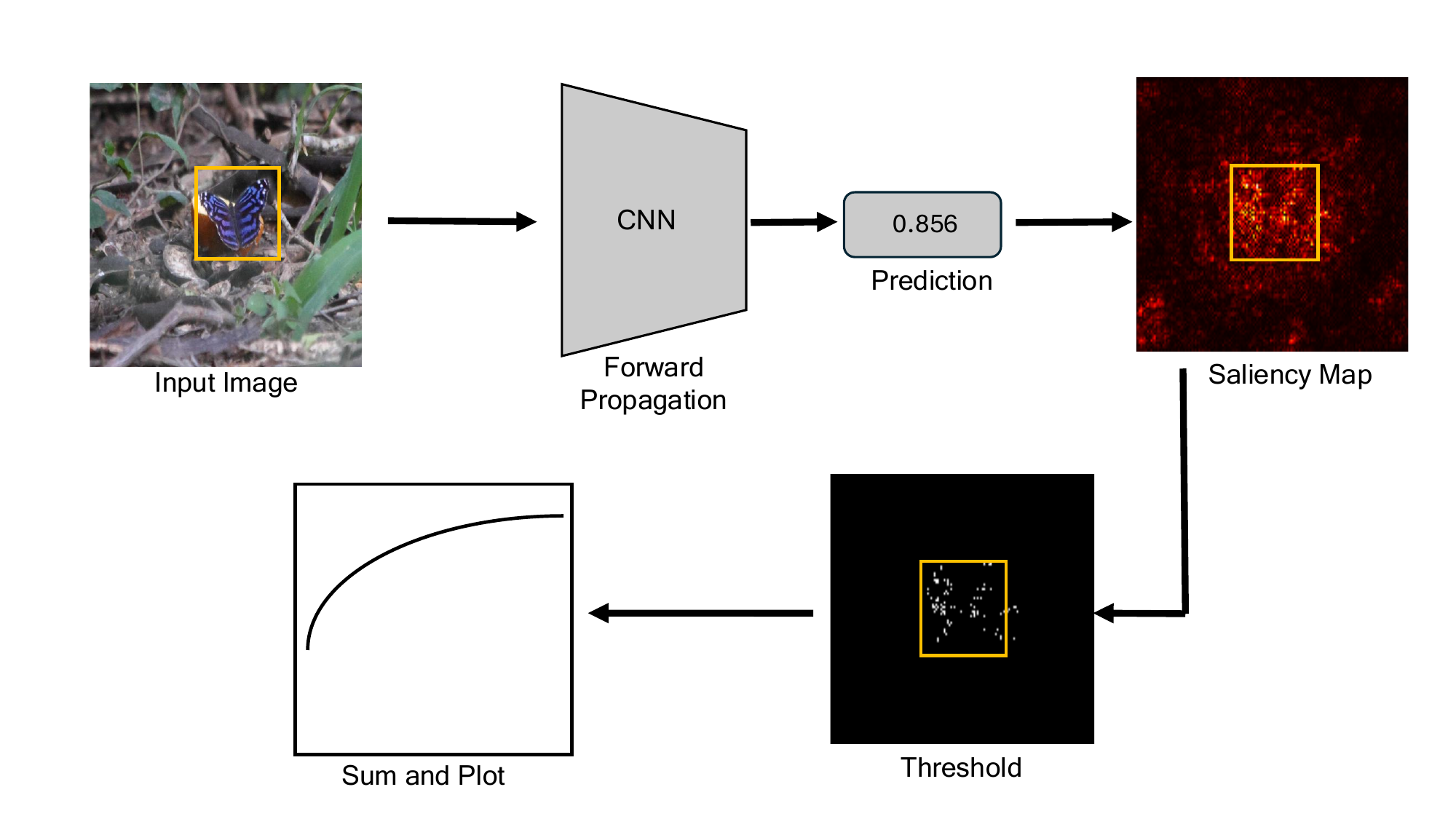}
    \caption{Simplified workflow for saliency map analysis showing the main stages. Inference yields a prediction and a corresponding saliency map, salient pixels are thresholded, creating a binary map. Pixels within the insect bounding box are counted and the portion of salient pixels is plotted for all thresholds and images. }
    \label{fig:sal-workflow}
\end{figure}

We implement a saliency map analysis pipeline. This process, visually shown in Figure \ref{fig:sal-workflow}, highlights how we measure the overlap between the model's most influential regions and the insect locations annotated by bounding boxes. Since our model has a single output neuron representing both classes, we focus only on positive gradients - those that signal the presence of insects (positive outputs). This is shown in Equation \ref{eqn1}, where $S$ is the saliency map, $I$ is the input image and $y$ is the output prediction. 

\begin{equation}\label{eqn1}
    S_k = \max\left(0, \frac{\partial y}{\partial I_k}\right)
\end{equation}

For each image, labelled by the index $k$, the resulting saliency map is normalised so that its values fall in range between 0 and 1. We then binarise these maps at different thresholds, $t$ (ranging from 0 to 1), this turns them into black-and-white images where only the most important pixels remain. This process is shown mathematically by Equation \ref{eqn2}:

\begin{equation}\label{eqn2}
S_{k,t} = 
\begin{cases} 
1, &  \text{if } S_k \geq t, \\
0, &  \text{if } S_k < t 
\end{cases}
\end{equation}

Next, to evaluate how well the saliency maps highlight insect image regions, we calculate $P^t_k$ as the proportion of pixels within a thresholded image, $S_{i, j}^{k, t}$, which lie within the range of any insect bounding boxes for that image, i.e. the proportion of salient pixels which are in the correct area. To gain an average performance across all images, we compute $\bar{P}^t$ as the average $P^t_k$ over all $k$ as shown by Equation \ref{eqn3}. We repeat this for all models and $\bar{P}$ is plotted over all values of $t$ to provide a visualisation of model performance. As the value of $t$ rises, $\bar{P}$ is expected to rise, indicating that the most salient pixels consistently lie within the image portion containing an insect, and not background regions which may be misleading. We plot $\bar{P}^t$ for our test data, producing a robust analysis of the learned “insect” representation. 

\begin{equation}\label{eqn3}
    \bar{P}_t = \frac{1}{N} \sum_{k=1}^{N} P_{k,t} 
\end{equation}


\subsection{Power}

Finally, minimisation of power consumption has been a key driver for the design decisions taken throughout this work and thus forms the final component of our analysis. To measure power consumption, we select a deployment platform, compile our models and execute them on the device. Firmware is written to target the ESP32-S3 chipset. We chose this platform as it is low-cost, readily available and supports the ESP-NN neural network acceleration library to speed up inference time. ESP-NN (available at: \url{https://github.com/espressif/esp-nn}) is a C++ library supported by the chip manufacturer, Espressive. It provides optimised neural network execution functions which are between 3 and 10 times faster than equivalent operations without using the library; in our firmware, we use version \texttt{1.0}. We also use the LILYGO® T-SIMCAM ESP32-S3 Development Board as a target. These are available for a retail price of around £20, including an embedded camera, wireless connectivity and the capacity to hold an SD and SIM card. Some alternatives could be boards which use chipsets that have support TensorFlow Lite builds (a list is available online: \url{https://ai.google.dev/edge/litert/microcontrollers/overview}), although, this paper does not intend to make specific hardware recommendations. 

Our testing firmware is very simple, images are captured at a resolution equal to the input size of the network, then, inference is then executed using the TensorFlow Lite Micro runtime \citep{tflite-micro} and this process is timed. We measure the time taken for inference and average over five inferences for each model. Current consumption is measured using a desk ammeter while idle and while running our programme. The system is powered using a bench power supply set to 3.3V.

\section{Results}\label{sec:results}
\begin{figure}[]
    \centering
    \includegraphics[width=\linewidth]{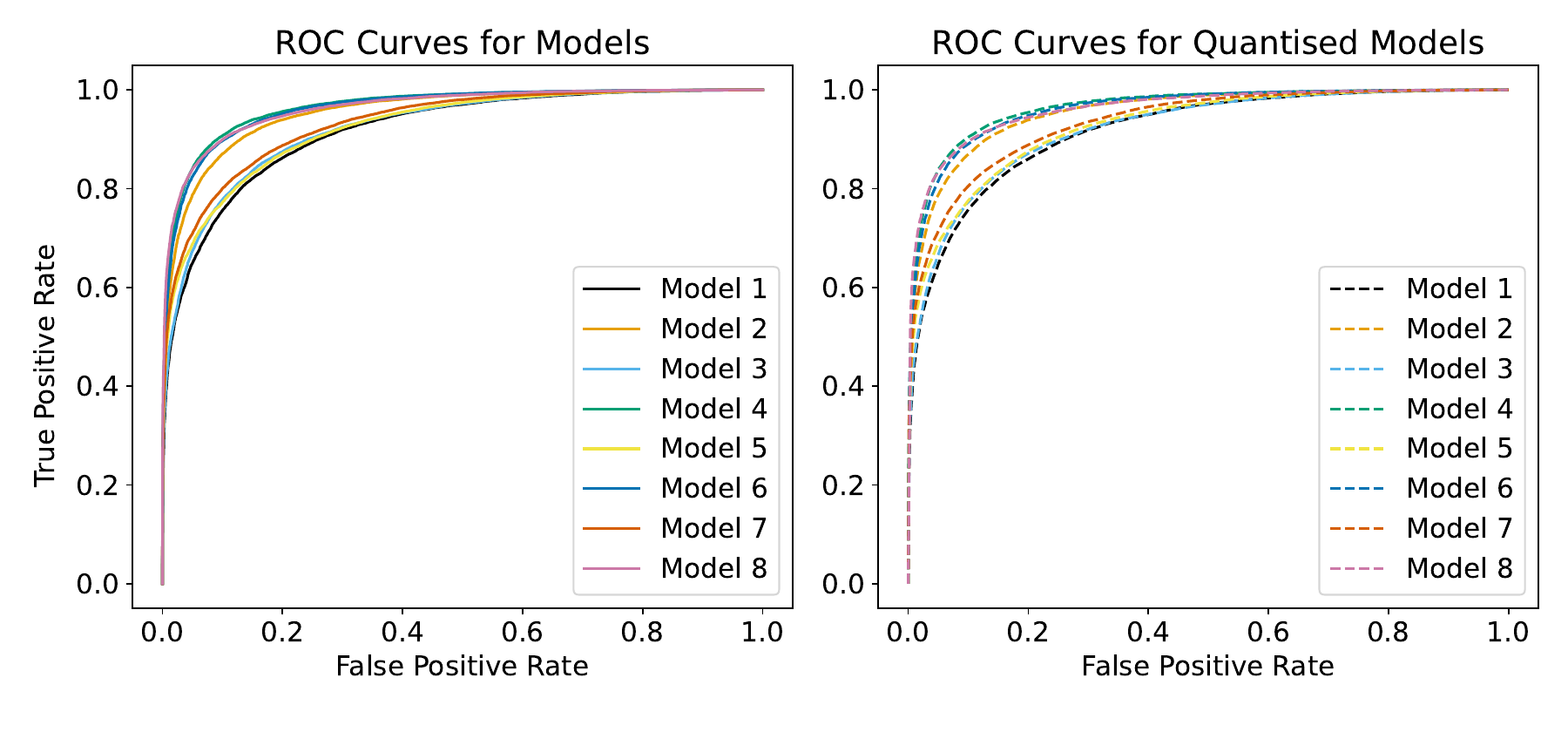}
    \caption{ROC curves for each model computed from the test dataset. Shows true positive rate (recall) against false positive rate (1 - specificity) over all thresholds. Original and quantised models 1-8 shown as solid and dotted lines respectively. }
    \label{fig:roc-curves}
\end{figure}

The test accuracy, precision, F1 scores and AUROC values for each model variant are summarised in Table \ref{tab:results}, and ROC curves are presented in Figure \ref{fig:roc-curves}. Further testing utilised the Insect-Detect, Ecostack, Pollinator-Detect and AMI-CT data with the tiling parameter, $n$, set to 2. Further F1 and precision scores for all datasets are available in the Supplementary Material (Table S1). Across all models, test accuracy ranges from 0.836 to 0.904, precision scores from 0.873 to 0.917 and F1 scores from 0.823 to 0.898 indicating strong performance. This trend is consistent across output thresholds, with test AUROC ranging from 0.918 to 0.964. We report quantised model accuracy figures in Table S1 and found that quantisation did not significantly affect these measures. The colourblind models generally underperform compared to models utilising RGB inputs. This is particularly evident for the more challenging Pollinator-Detect dataset, where AUROC values fall considerably.  

On the field datasets, model performance is mixed. The out-of-domain Pollinator-Detect dataset proved the most challenging across all models and performance for the smaller models is especially weak for these examples, with all models performing poorly. Performance on Insect-Detect and Ecostack is strong and consistently outperforms the test set. This result is not surprising as examples from these datasets are included in the training mix (Table \ref{tab:data}) and have less diverse backgrounds than those in the iNat2017 subset. On the AMI-CT dataset, out-of-domain performance remains strong. The results indicate that our models demonstrate high discriminative power for unseen data with insects on simple, uncluttered backgrounds and strong overall performance on data from the same distribution as the training source.


\begin{table}[!htp]
\resizebox{\textwidth}{!}{%
\begin{tabular}{lcccccccc}
Model &
  \begin{tabular}[c]{@{}c@{}}Accuracy \\Test \\ \end{tabular} &
  \begin{tabular}[c]{@{}c@{}}Precision \\Test \\ \end{tabular} &
  \begin{tabular}[c]{@{}c@{}}F1 \\ Test \\ \end{tabular} &
  \begin{tabular}[c]{@{}c@{}}AUROC \\ Test \\ \end{tabular} &
  \begin{tabular}[c]{@{}c@{}}AUROC \\Insect-Detect\\ n=2\end{tabular} &
  \begin{tabular}[c]{@{}c@{}}AUROC \\Ecostack\\ n=2\end{tabular} &
  \begin{tabular}[c]{@{}c@{}}AUROC \\Pollinator-Detect\\ n=2\end{tabular} &
  \begin{tabular}[c]{@{}c@{}}AUROC \\AMI-CT\\ n=2\end{tabular} \\
  \hline
\begin{tabular}[c]{@{}l@{}}1\end{tabular} &
  0.836 &
  0.873 &
  0.823 &
  \begin{tabular}[c]{@{}c@{}}0.918\end{tabular} &
  \begin{tabular}[c]{@{}c@{}}0.955\end{tabular} &
  \begin{tabular}[c]{@{}c@{}}0.840\end{tabular} &
  \begin{tabular}[c]{@{}c@{}}0.635\end{tabular} &
  \begin{tabular}[c]{@{}c@{}}0.818\end{tabular} \\
\begin{tabular}[c]{@{}l@{}}2\end{tabular} &
  0.883 &
  0.874 &
  0.887 &
  \begin{tabular}[c]{@{}c@{}}0.955\end{tabular} &
  \begin{tabular}[c]{@{}c@{}}0.959\end{tabular} &
  \begin{tabular}[c]{@{}c@{}}0.887\end{tabular} &
  \begin{tabular}[c]{@{}c@{}}0.588\end{tabular} &
  \begin{tabular}[c]{@{}c@{}}0.841\end{tabular} \\
\begin{tabular}[c]{@{}l@{}}3\end{tabular} &
  0.817 &
  0.768 &
  0.835 &
  \begin{tabular}[c]{@{}c@{}}0.923\end{tabular} &
  \begin{tabular}[c]{@{}c@{}}0.970\end{tabular} &
  \begin{tabular}[c]{@{}c@{}}0.841\end{tabular} &
  \begin{tabular}[c]{@{}c@{}}0.605\end{tabular} &
  \begin{tabular}[c]{@{}c@{}}0.854\end{tabular} \\
\begin{tabular}[c]{@{}l@{}}4\end{tabular} &
  0.900 &
  0.940 &
  0.894 &
  \begin{tabular}[c]{@{}c@{}}0.966\end{tabular} &
  \begin{tabular}[c]{@{}c@{}}0.972\end{tabular} &
  \begin{tabular}[c]{@{}c@{}}0.941\end{tabular} &
  \begin{tabular}[c]{@{}c@{}}0.637\end{tabular} &
  \begin{tabular}[c]{@{}c@{}}0.859\end{tabular} \\
\begin{tabular}[c]{@{}l@{}}5\end{tabular} &
  0.844 &
  0.862 &
  0.836 &
  \begin{tabular}[c]{@{}c@{}}0.925\end{tabular} &
  \begin{tabular}[c]{@{}c@{}}0.976\end{tabular} &
  \begin{tabular}[c]{@{}c@{}}0.868\end{tabular} &
  \begin{tabular}[c]{@{}c@{}}0.595\end{tabular} &
  \begin{tabular}[c]{@{}c@{}}0.859\end{tabular} \\
\begin{tabular}[c]{@{}l@{}}6\end{tabular} &
  0.896 &
  0.908 &
  0.898 &
  \begin{tabular}[c]{@{}c@{}}0.963\end{tabular} &
  \begin{tabular}[c]{@{}c@{}}0.974\end{tabular} &
  \begin{tabular}[c]{@{}c@{}}0.932\end{tabular} &
  \begin{tabular}[c]{@{}c@{}}0.691\end{tabular} &
  \begin{tabular}[c]{@{}c@{}}0.835\end{tabular} \\
\begin{tabular}[c]{@{}l@{}}7\end{tabular} &
  0.857 &
  0.866 &
  0.849 & 
  \begin{tabular}[c]{@{}c@{}}0.934\end{tabular} &
  \begin{tabular}[c]{@{}c@{}}0.982\end{tabular} &
  \begin{tabular}[c]{@{}c@{}}0.866\end{tabular} &
  \begin{tabular}[c]{@{}c@{}}0.600\end{tabular} &
  \begin{tabular}[c]{@{}c@{}}0.860\end{tabular} \\
\begin{tabular}[c]{@{}l@{}}8\end{tabular} &
  0.904 &
  0.917 & 
  0.898 &
  \begin{tabular}[c]{@{}c@{}}0.964\end{tabular} &
  \begin{tabular}[c]{@{}c@{}}0.992\end{tabular} &
  \begin{tabular}[c]{@{}c@{}}0.959\end{tabular} &
  \begin{tabular}[c]{@{}c@{}}0.680\end{tabular} &
  \begin{tabular}[c]{@{}c@{}}0.872\end{tabular}
\end{tabular}%
}
\caption{Table of numerical results for each model showing test accuracy, precision, F1 scores and area under the ROC curve for respective datasets. Insect-Detect, Ecostack, Pollinator-Detect and AMI-CT are tiled such that the resulting images are now one quarter of the size of the original images; their tiling parameter, n, is set to 2. }
\label{tab:results}
\end{table}

\begin{figure}[!htp]
    \centering
    \includegraphics[width=\linewidth]{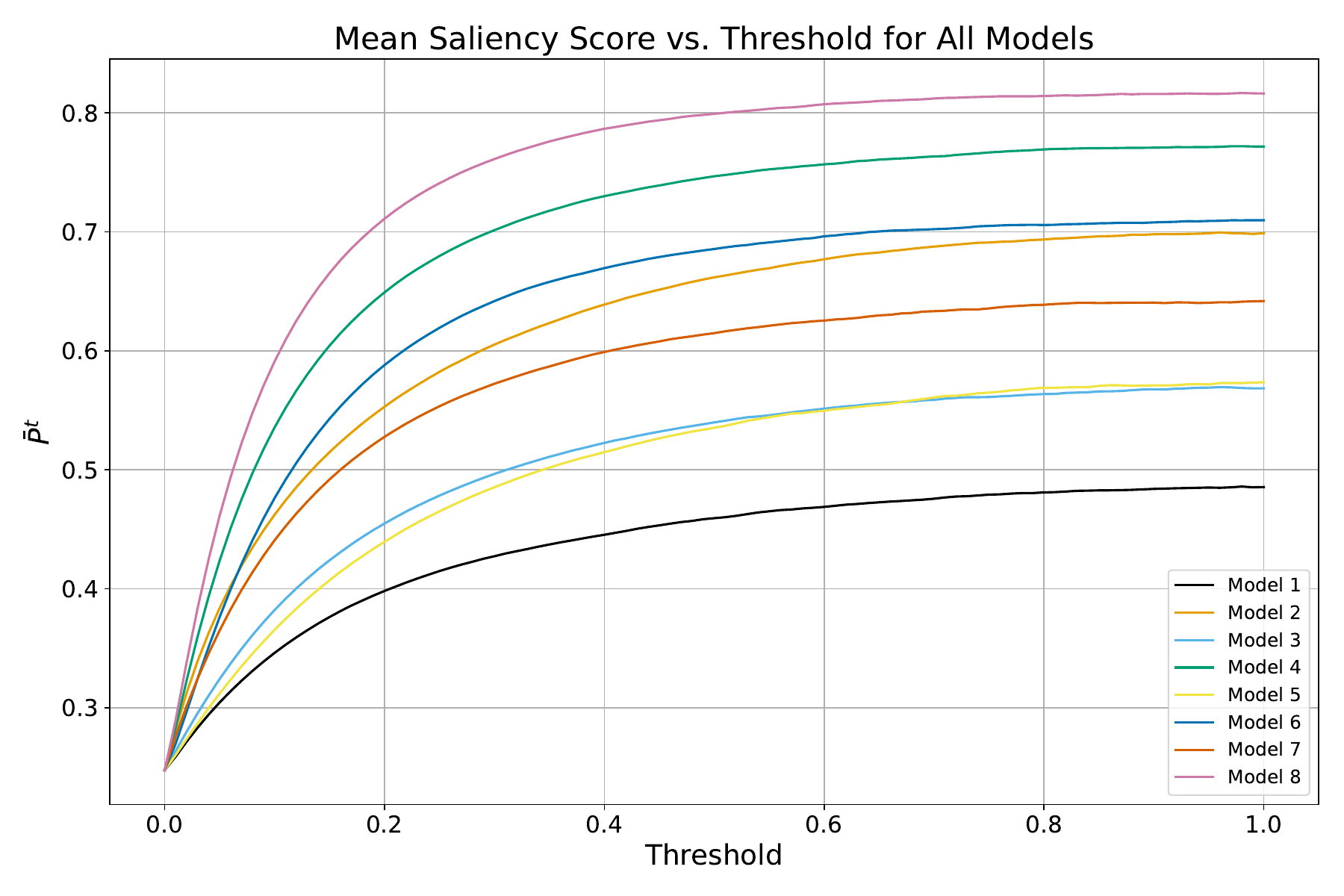}
    \caption{Resultant curves for our saliency score, $\bar{P}$ for each threshold, $t$, averaged over each image our test dataset measuring the alignment of saliency maps with insect image regions. Each solid line depicts models 1-8.}
    \label{fig:pt_curves}
\end{figure}
\begin{figure}[H]
    \centering
    \includegraphics[width=\linewidth]{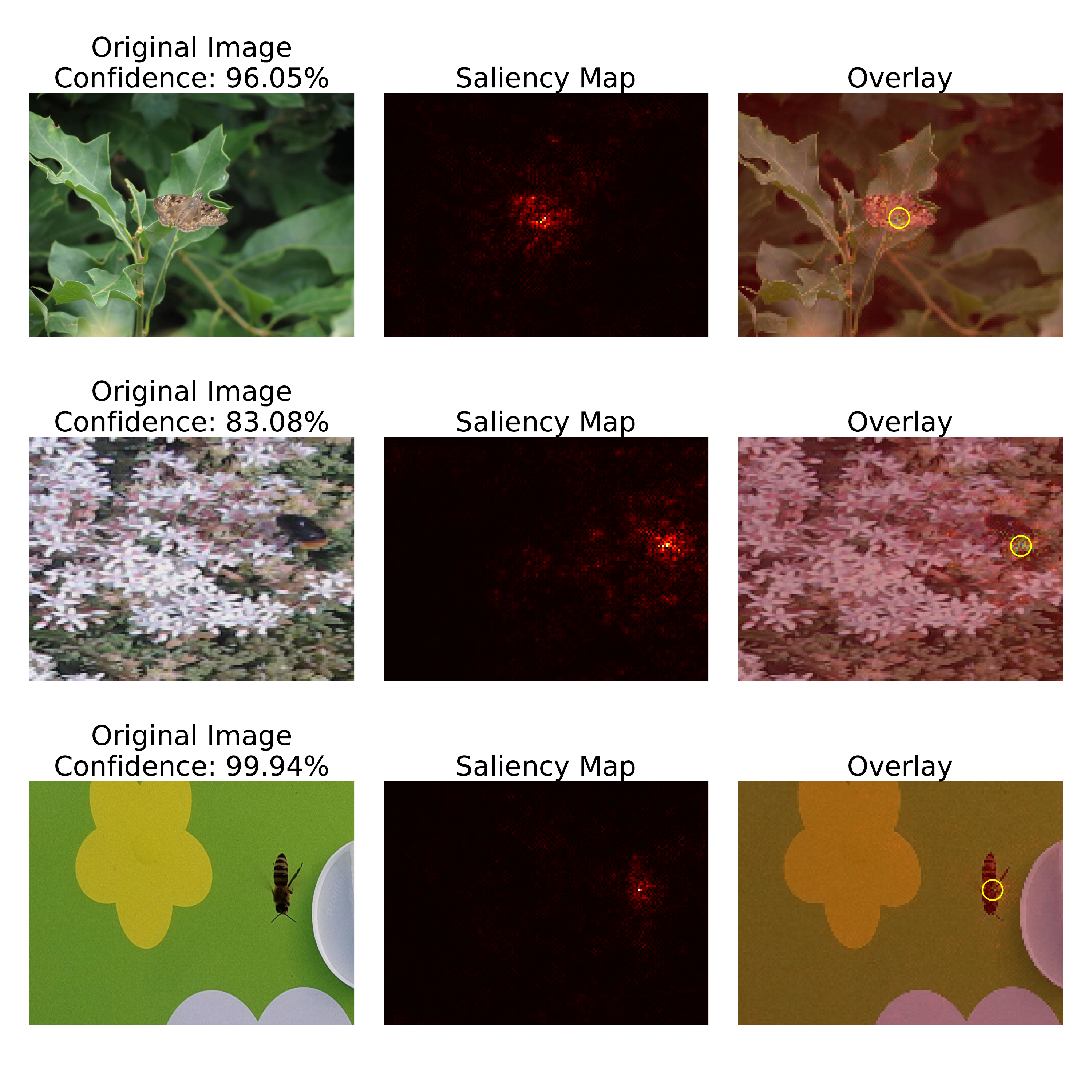}
    \caption{Example predictions from our test dataset, executed using model 8. Input images with their prediction confidence are shown in the leftmost column, saliency maps are shown in the center column and the rightmost column overlays this with the input image. In the rightmost column, the most salient pixel is highlighted with an encompassing circle. The image on the first depicts a butterfly from the iNaturalist 2017 challenge dataset \citep{inaturalist}. The image on the second row depicts a bee in flight from the Ecostack dataset \citep{ecostack}. The image on the final row depicts a bee in flight from the Insect-Detect dataset \citep{Insect-detect}. The final 2 rows are examples tiled image regions where they are cropped four times smaller than the original dataset image. }
    \label{fig:in-dom}
\end{figure}
\begin{figure}[H]
    \centering
    \includegraphics[width=\linewidth]{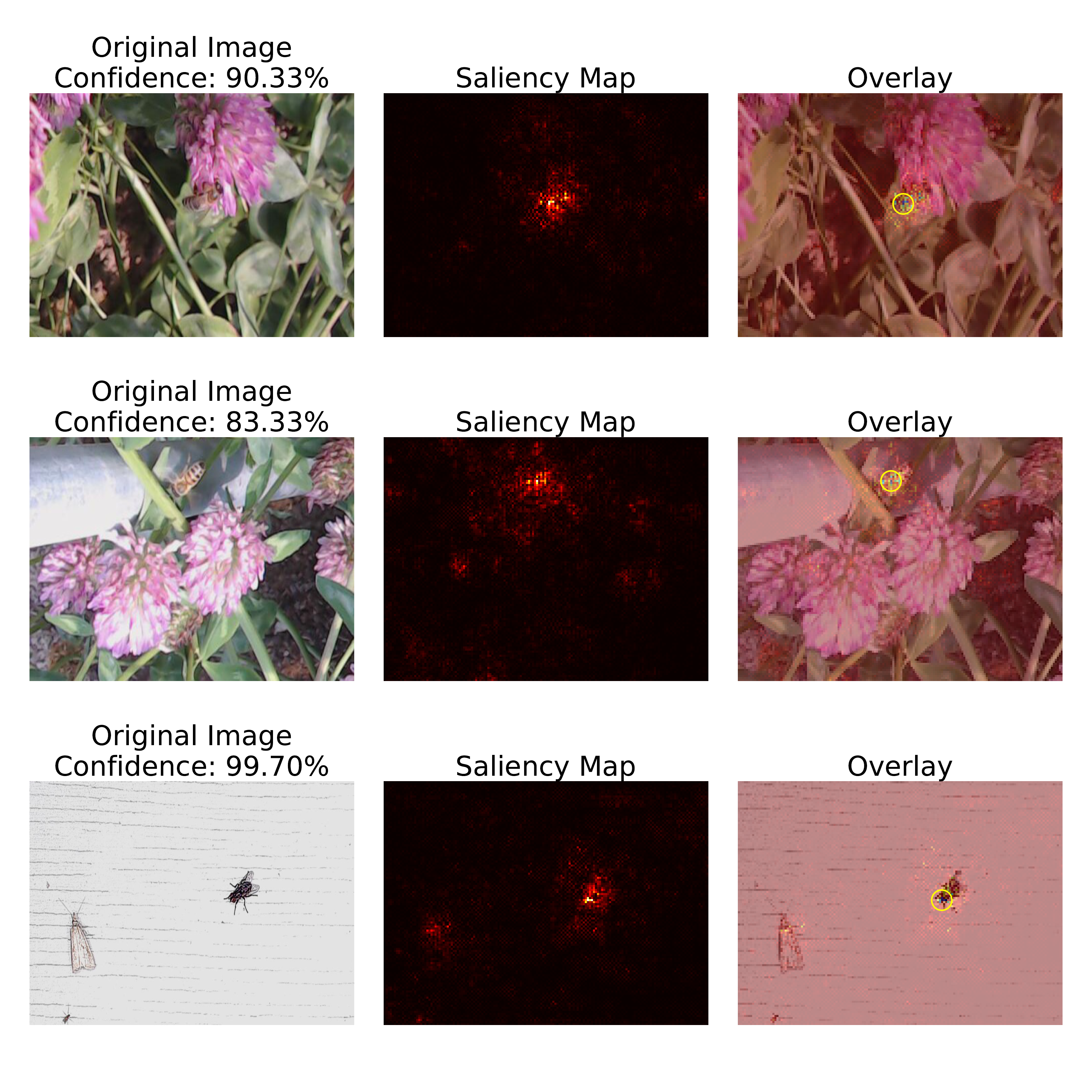}
    \caption{Example predictions from datasets unseen in training, executed using model 8. Input images with their prediction confidence are shown in the leftmost column, saliency maps are shown in the centre column and the rightmost column overlays this with the input image. In the rightmost column, the most salient pixel is highlighted with an encompassing circle. The first and second rows depict insect images from the Pollinator-Detect dataset \citep{pollinator-detect} In the second row, the insect is in flight. The final row depicts two insects in an AMI-CT image \citep{ami-dataset}.}
    \label{fig:out-dom}
\end{figure}

Curves for our saliency map analysis are presented in Figure \ref{fig:pt_curves}. These curves reveal a consistent pattern with the accuracy analysis, the colourblind models display a less trustworthy learned representation, as evidenced by a low proportion of salient pixels within insect bounding boxes. In contrast, the RGB models consistently show that the most salient pixels lie in areas of the images belonging to insects, suggesting a more robust learned representation which can be more easily generalised; this is also backed up by RGB models performing better on out-of-domain examples. Examples of saliency map predictions are shown for in-domain data in Figure \ref{fig:in-dom} and for out-of-domain data in Figure \ref{fig:out-dom}.

\begin{figure}[H]
    \centering
    \includegraphics[width=\linewidth]{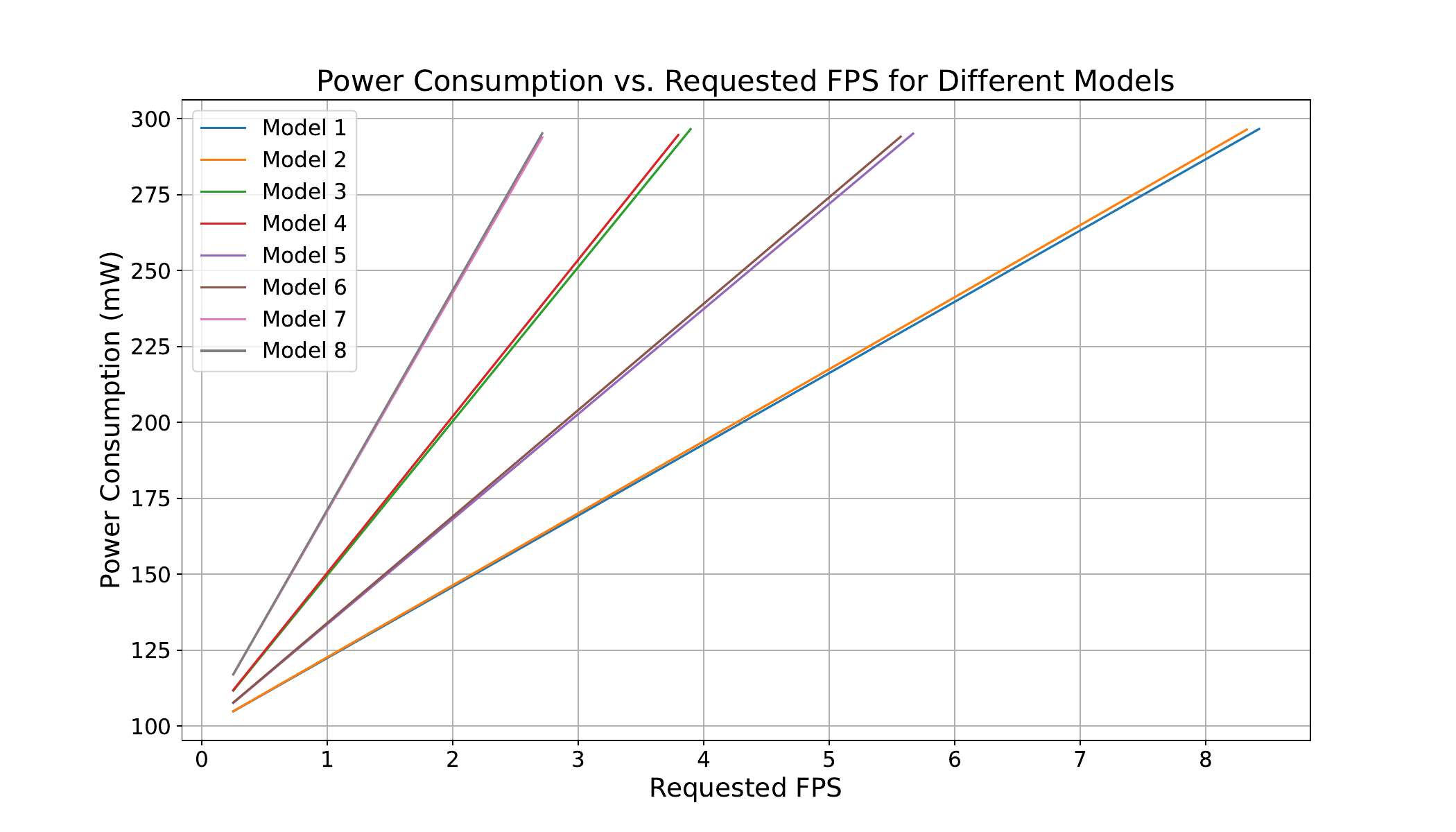}
    \caption{Average power consumption for each model versus operational frames per second processed by the ESP32-S3 chipset.}
    \label{fig:power}
\end{figure}

Power consumption and inference latency characteristics of each model are shown in Figure \ref{fig:power}. This also shows how the number of frames-per-second processed by the system can affect power consumption as this defines the proportion of time the MCU executes instructions vs the proportion of time it is sitting idle. Operating at their highest throughput, all models consume less than 300mW and at half throughput this can be reduced to around 185mW. The most lightweight models, 1 and 2, can process over eight frames per second. The most accurate model, 8, can process over two frames per second. 

\section{Discussion}\label{sec:discussion}
Our methodology balances the complex requirements of an ideal camera trap trigger, offering a novel approach to detecting ectotherms in camera trap systems while adhering to the data volume and power constraints of remote environments. We have diverged from the traditional approach of PIR sensors and the popular insect camera trap approach of object detection on-device to instead offer lightweight binary classification architectures suitable for MCUs. Below, we provide further insights from our results, how our method compares with existing “fully-fledged” insect camera traps (meaning those deployed in the field from prior publications) and provide suggestions how others may integrate this method into future designs. We also discuss the limitations of our method and point to some directions for future work. 

\subsection{Performance}

Our results demonstrate that the models are excellent candidates for insect triggers, achieving high accuracy, precision, and F1 scores, with strong specificity and recall across output thresholds, even on AMT-CT which was unseen in training. Saliency map analysis further validated model reliability. Colorblind models relied more often on spurious background information, as shown by a significant drop in $\bar{P}_t$. In contrast, RGB scored higher $\bar{P}_t$, boosting confidence in their performance. Importantly, our approach operates effectively on realistic microcontroller platforms, maintaining low power consumption. Overall, our methodology delivers on each of the aspects of a successful camera trap trigger identified. 

Although this work does not provide a fully-fledged insect monitoring system, we compare the characteristics of our hardware test with those developed by others in Table \ref{tab:look-at-me}. This reveals several advantages: (1) the power consumption is low enough to facilitate multiple-week deployments given a moderate battery size of 110-Wh, this is far longer than existing systems; (2) We showcase a new hardware platform and AI methodology that has not yet been used in insect monitoring; (3) The framerate achieved by our model execution is greater than the only other MCU-based hardware in the list (\citep{darras}, row 2) and considerably lower framerates than devices deployed with more capable computing hardware. Overall, this niche potentiates large power consumption benefits at lower framerates.

\begin{table}[H]
\begin{tabular}{llllll}
System    & \begin{tabular}[c]{@{}l@{}}Target\\ Hardware\\ Target\end{tabular} & \begin{tabular}[c]{@{}l@{}}Power\\ Consumption \\(W)  \end{tabular} & \begin{tabular}[c]{@{}l@{}}Frames\\  per \\ second\end{tabular} & \begin{tabular}[c]{@{}l@{}}Battery \\life \\(days)\end{tabular} & \begin{tabular}[c]{@{}l@{}}Computer vision \\ method on device\end{tabular}                  \\
\hline
\citep{pict}      & RPi Zero W                                                         & 0.62 - 1.13                                                                                & 10-90                                                           & $\sim$7.4-4                                                                             & None                                                                                         \\
\citep{darras}    & OpenMV H7+                                                         & 1.3                                                                                        & \begin{tabular}[c]{@{}l@{}}0.41-\\ 1.75\end{tabular}            & $\sim$3.5                                                                               & \begin{tabular}[c]{@{}l@{}}Object detection,\\ Blob detection,\\ Classification\end{tabular} \\
\citep{Insect-detect} & \begin{tabular}[c]{@{}l@{}}RPi Zero 2 W\\  + OAK-1\end{tabular}    & 4.4                                                                                        & 12.5                                                            & $\sim$1                                                                                 & Object detection                                                                             \\
\citep{amt} & RPi 4B                                                             & 12.5W                                                                                      & 0.5-2                                                           & 0.37                                                                                    & \begin{tabular}[c]{@{}l@{}}Blob Detection, \\ Classification\end{tabular}                    \\
\citep{bjerge-jetson} & \begin{tabular}[c]{@{}l@{}}NVIDIA \\ Jetson Nano\end{tabular}      & 5-8W                                                                                       & 0.33                                                            & 0.92-0.57                                                                               & Object Detection                                                                             \\
\citep{motion-vectors}   & RPi 4B                                                             & 3.75-6.25                                                                                  & 20                                                              & $\sim$1.2-0.73                                                                          & \begin{tabular}[c]{@{}l@{}}Video Motion Detection,\\ Object Detection\end{tabular}           \\
Ours      & ESP32-S3                                                            & 0.125-0.3                                                                                  & 0.3-8                                                           & 36.7-15.3                                                                               & Lightweight Binary Classifiers                                                              
\end{tabular}
\caption{Comparison of existing systems with our hardware test. All statistics lifted from respective publications. Power consumption figures are taken from each paper quoted without any additional lighting methods for night-time operation. Battery life is computed as a product of power consumption given a 110-Wh battery. }
\label{tab:look-at-me}
\end{table}

\subsection{Future Directions and Applications}

Advancements in low-power AI and microcontrollers are rapidly expanding the potential for scalable biodiversity monitoring. As AI models become more sophisticated and edge devices become cheaper and more energy-efficient monitoring devices will benefit from improved performance, reduced power consumption and enhanced accessibility. Progress in camera trapping hardware has already demonstrated some of these gains \citep{darras} and continued improvements in model design, data accessibility and AI methodologies are expected to push this further. Moving forward, it will be essential to focus on both the scalability of these technologies and the trustworthiness of AI models to ensure their reliability and adoption in ecological monitoring.

This is especially crucial for studying insect declines in tropical regions, where data are scarce and conventional monitoring methods are impractical. For example, monitoring insects in shaded or confined ecosystems—such as beneath dense jungle canopies or within treetops—remains difficult due to limited access \citep{moth-pir}. A lightweight, efficient device will transform capability as alternatives require ample physical space and continuous sunlight to facilitate long deployments. 

Additionally, MCUs could be adapted to be a standalone trigger system. Here, the MCU (including its camera), would operate independently to the main camera system and output a signal to trigger it. For camera traps with low idle power consumption --many of which have external trigger inputs-- this setup could offer significant energy savings. However, this approach would introduce the trigger latency of the camera trap so such a system may be probative for fast-moving subjects. Thus, we suggest this could be viable in cases where insects remain stationary for a period such as pollinator monitoring or systems using light traps/lures.

\subsection{Limitations and Future Work}

Although promising, this method presents several limitations that warrant future work. Notably, the small input size of our models has been necessary to enable lightweight computation; however, the resulting low input resolution restricts the field of view or requires tiling. Emerging advancements in lightweight shallow convolutional networks offer the potential for larger input sizes without exceeding the same compute budget \citep{fomo}. Incorporating these models could enable the processing of higher-resolution imagery. 

The framerate of our MCU test is considerably lower than competing systems that record video continuously. This may limit applications requiring accurate tracking of insect movement, this is crucial, as continuous tracking has been shown to provide a better measure of occupancy, avoiding multiple detections per visit \citep{bjerge-jetson}. Additionally, low framerates increase the risk of missing fast-moving insects that quickly pass through the camera's field of view. This constraint, driven by model execution latency, can be addressed through strategies such as distributing inferences across frames \citep{motion-vectors} and applying further compression techniques such as network pruning \citep{pruning}. 

We also note that although power consumption is markedly lower than similar systems, it cannot rival the idle consumption of a traditional PIR trigger, which typically consumes <<1mW. It is clear that the further reduction of energy will continue to provide scalability benefits. We suggest further firmware engineering aspects be used, such as MCU sleep modes and under-clocking. More computationally efficient models will assist this by allowing the system to enter low-power states more frequently, between model runs. 

Our models have also struggled generalising to the out-of-domain Pollinator-Detect dataset, suggesting they require refinements for some downstream use cases. This could be addressed in the selection of training data. As more field datasets become available triggers can be trained on a more diverse mix. The method we have shown allows further training data to be integrated easily. 

Finally, while in this work we focus on addressing insect monitoring, we stress that these concepts could be applied to the retrieval of animal images more generally. \citeauthor{meek} found that faster trigger times are one of the most common requests from conventional camera trap users \citep{meek} and PIR sensors have been shown to generate many false triggers; the Snapshot Serengeti camera trap trail data contains over 1.2 million images and only 26\%  were found to contain animals \citep{snapshotser}. Our method, with zero latency, and high accuracy provides new directions. This method as a replacement for, or working alongside PIR triggers in traditional camera traps may help to address low recall of fast-moving small animals and the large number of false positive triggered images that hinder camera trap scalability. 

\section{Conclusion}\label{sec:conclusion}
Using the latest innovations in MCU hardware and ultra-lightweight supervised learning, we have shown how insect monitoring can be made scalable by addressing the challenging trigger problem. Through extensive analysis, we have established trust in our methods, showing our models are well-suited for this purpose. The demonstration of our models deployed on hardware saves energy using readily available components. Our models offer improvements in scalability, aiding future systems to be deployed in greater quantity, improving sampling density without the need for excessive manual labour. As such methods are essential for the timely assessment of biodiversity trends, we have provided tools and guidance to encourage others towards this direction.  

\bibliographystyle{unsrtnat} 


\end{document}